\documentclass[twocolumn]{aastex631}
\usepackage{graphicx}
\usepackage{amsmath}
\usepackage{natbib}
\usepackage[T1]{fontenc}
\usepackage{float}

\shorttitle{The Geometric Crisis in Cygnus X-3}
\shortauthors{White}

\begin{document}

\title{The Geometric Crisis in Cygnus~X-3: Limitations of Wind-Fed Accretion and the Case for Roche-Lobe Overflow}

\author{Nicholas E. White}
\affiliation{Department of Physics, George Washington University \\
Corcoran Hall, 725 21st Street NW \\
Washington, DC 20052, USA}
\email{newhite@gwu.edu}

\begin{abstract}
Cygnus~X-3 is a Galactic X-ray binary with a 4.8-hr orbital period operating in the
ultraluminous regime. Although the system is viewed at relatively low inclination
($i\approx28^\circ$), it exhibits a deep orbital modulation. Recent \textit{IXPE}
observations show strong linear polarization orthogonal to the radio jet, indicating that
the X-ray emission is dominated by reflection from the inner walls of a supercritical
\textbf{outflow} funnel.

We propose a ``Hybrid'' Roche-lobe overflow (RLOF) scenario in which a massive Wolf--Rayet donor
effectively fills its Roche lobe with a focused wind driving a super-Eddington accretion
stream. Using a numerical synthesis of the folded light curve, we show
that the modulation is reproduced when the central funnel is periodically occulted by a
vertically extended, shock-heated ``Turbulent Wall'' formed by stream impact on the outer
disk rim. This produces a phase lag ($\Delta\phi\approx0.11$) between X-ray minimum and
binary conjunction, with extended attenuation by the WR wind defining a broader ``Suppression
Region''.

This geometry explains the enhanced iron-line equivalent width during X-ray minimum via a
coronagraphic effect. The large radial-velocity amplitude of Fe\,{\sc xxvi} measured by
\textit{XRISM} ($K_{\rm obs}\approx430$~km~s$^{-1}$) and its zero-crossing at $\phi_X=0.0$
arise naturally in the stream-impact region rather than from orbital motion of the compact
object.

Finally, we show that the observed secular orbital expansion ($\dot P>0$) follows directly
from highly non-conservative mass transfer with inner-disk mass loss, indicating that
Cygnus~X-3 is a stable, long-lived system in a supercritical accretion regime.
\par\medskip
\noindent\textbf{Note:} This version includes an erratum
accepted by ApJ, appended on pages 18--19.
\end{abstract}
\keywords{X-rays: binaries --- stars: Wolf--Rayet --- accretion, accretion disks --- polarization}

\section{Introduction} \label{sec:intro}
Since its discovery in 1966 \citep{Giacconi1967}, Cygnus X-3 has remained one of the most enigmatic objects in the Galaxy. It is the only known Galactic High-Mass X-ray Binary (HMXB) comprising a Wolf--Rayet (WR) star and a compact object, serving as the definitive local analog for Ultraluminous X-ray Sources (ULXs). Defined by its smooth, quasi-sinusoidal 4.8-hour orbital modulation \citep{Parsignault1972}, the system alternates between radio-quiet and massive flaring states \citep{Waltman1994}. Spectroscopically, it is dominated by the ``Iron Forest,'' the strongest Fe K emission line complex of any X-ray binary \citep{Serlemitsos1975}. While early observations could only resolve a single broad feature, modern instruments (\textit{ASCA}, \textit{Chandra}, \textit{XRISM}) have revealed this complex structure in detail \citep{Kitamoto1994, Kallman2019, XRISM2024}.

The current consensus for the geometry and accretion dynamics of Cygnus X-3 is that a compact object (black hole or neutron star) orbits inside the dense stellar wind of its WR companion \citep{vanKerkwijk1996} and is powered by classical Bondi--Hoyle--Lyttleton (BHL) wind accretion \citep{Hoyle1939, Bondi1944, Pringle1974, Szostek2008, Antokhin2022}. The 4.8-hour modulation is caused by varying path lengths through the wind to the X-ray source \citep{Pringle1974, Szostek2008}. Detailed modeling by e.g., \citet{Antokhin2022} has shown that wind models generate the orbital modulation depth and shape, but require substantial optical depth variations of order unity. 

A challenge for this model is that it predicts orbital-phase–dependent wind absorption. However, repeated X-ray spectral measurements across all source states show no corresponding variation in the inferred column density; instead, the absorber is observed to be effectively ``grey'' \citep{Zdziarski2012}. While Thomson scattering has been invoked to explain such energy-independent attenuation, this interpretation would require a wind that is uniformly and highly ionized along the line of sight. High-resolution X-ray spectroscopy with \textit{Chandra} HETG demonstrates that the Wolf--Rayet wind in Cyg~X-3 is instead strongly ionization-stratified, retaining significant populations of bound ions \citep{Kallman2019}, a conclusion now independently reinforced by the latest \textit{XRISM} observations \citep{XRISM2024}. These observations therefore challenge the assumption of a uniformly, fully ionized wind required by this explanation.

A recent observational development has provided critical new insight into the accretion geometry of Cygnus X-3. The Imaging X-ray Polarimetry Explorer (\textit{IXPE})\citep{Weisskopf2022} has revealed a remarkably high degree of linear polarization ($PD \approx 20\%$) in the hard state, oriented orthogonal to the radio jet \citep{Veledina2024a}. This signal implies the emission is dominated by reflection off the inner walls of a super-critical outflow funnel viewed at low inclination ($i \approx 28^{\circ}$) \citep{Veledina2024a}. Notably, the polarization degree exhibits a distinct dip at 6.4 keV (neutral Iron K$\alpha$), indicating dilution by isotropic unpolarized fluorescence \citep{Mikusincova2025}. \citet{Veledina2024a} suggest this is an ultra-luminous X-ray (ULX) source located in our Galaxy, where we observe the interior of the funnel, with the central source still hidden from view. This ULX interpretation means the Cygnus X-3 is a local example of ``slim disks'' or ``puffed-up'' accretion flows \citep{Abramowicz1988}, where radiation pressure inflates the disk into a geometrically thick torus ($H/R \sim 1$) with a central funnel \citep{Ohsuga2011, Jiang2014}. As we will show here, this new discovery presents further challenges to the wind accretion model and instead favors an alternate ``Occultation Model'' proposed by \citet{White1982}, which argued that the orbital modulation is geometric, primarily caused by a thickened bulge at the edge of the accretion disk.

\section{Accretion Geometry} 
\label{sec:models_mass}

The fundamental conflict facing current models is the requirement for a low inclination angle. Under the historic assumption of $i \approx 60^\circ$, the binary mass function implied a low-mass companion ($M_{WR} \lesssim 5M_\odot$) unable to fill its Roche-lobe. The \textit{IXPE} results fundamentally alter this calculus by showing this is a low inclination system. The inclination, $i$, was first estimated at $\approx 25^\circ$ \citep{Veledina2024a} and subsequently refined to the range $26^{\circ}\text{--}28^{\circ}$ by \citet{Veledina2024b}. Historically, the mass function derived from infrared lines of the WR wind was small ($f(M) \approx 0.027 M_{\odot}$; \citealt{Hanson2000}). At high inclinations, this allowed for a low-mass donor. However, combining the low \textit{IXPE} inclination with the standard assumption that the compact object is a black hole ($M_X \gtrsim 5\text{--}10 M_{\odot}$) forces a radical re-evaluation. To satisfy the mass function at $i \approx 27^{\circ}$, the companion must be a massive Wolf--Rayet star ($M_{WR} \gtrsim 15\text{--}20 M_{\odot}$). This high-mass scenario is independently suggested by recent high-resolution spectroscopy from \textit{XRISM}, which detects a large velocity amplitude of $K_X \sim 430$ km s$^{-1}$ \citep{Miura2025}. While we argue in Section \ref{sec:kinematics} that this feature traces the accretion stream dynamics rather than pure orbital motion, its extreme magnitude is physically difficult to reconcile with the low-velocity flows expected in a low-mass, wind-fed system. Unlike the low-mass scenario, such a massive WN7-type progenitor has an extended photospheric radius fully capable of filling its Roche-lobe (e.g., \citealt{Hamann2006, Grafener2012}).

This geometry resurrects the model of \citet{White1982}, who argued that the modulation is caused by a thickened bulge or wall at the edge of the accretion disk which occults the Accretion Disk Corona (ADC). This obscuration geometry identifies Cygnus X-3 as a super-critical analog to the ADC and dipping Low-Mass X-ray Binaries (LMXBs) \citep{White1982, White1995, DiazTrigo2006}. While \citet{White1982} envisioned a hydrostatic ADC typical of LMXBs, the super-Eddington nature of Cygnus X-3 implies the scattering surface is the inner skin of a radiation-driven funnel \citep{Poutanen2007, Veledina2024a}. However, the geometric principle---occultation of an extended scattering source by a disk bulge---remains identical. 

We refer to this as a ``Hybrid'' Roche-lobe overflow (RLOF) scenario, as it combines the intense stellar wind characteristic of a WR star with the focused accretion stream of RLOF. 

%
%
\section{Modeling the Orbital Lightcurve}
\label{sec:modeling}

To test the physical validity of the ``Hybrid'' RLOF model, we developed a simple geometric model that synthesizes the orbital lightcurve $L(\phi)$ as the product of two distinct modulation components: (1) geometric occultation of an extended X-ray source by a variable disk rim, and (2) phase-dependent absorption by the Wolf--Rayet stellar wind. Critically, to reproduce the smooth morphology of the eclipse, the central X-ray source is modeled not as a point, but as an extended scattering photosphere filling the inner funnel. 

The \textit{IXPE} measurements provide two boundary conditions for this model. First, the detection of high linear polarization orthogonal to the radio jet axis constrains the binary inclination to a low value. \citet{Veledina2024b} derive a strict range of \begin{equation} 26^\circ < i < 28^\circ \end{equation} $-$ we adopt $28^\circ$. Second, the polarization degree ($\sim 20\%$) and angle in the hard state imply that the emission is dominated by reflection from the walls of a collimated funnel. \citet{Veledina2024a} infer a funnel half-opening angle of \begin{equation} \theta_{funnel} \lesssim 15^\circ \end{equation} measured from the polar axis. Geometrically, this requires the X-ray source to be extended rather than compact. A source filling this funnel would have a full angular width of $\sim 30^\circ$. This constraint provides the physical upper bound for the \textbf{extended source} parameter used in our numerical modeling.

The orbital modulation of Cygnus X-3 is remarkably stable over decades. To model the system's geometry, we utilize the high-precision folded X-ray light curve presented by \citet{Choudhury2004}. While the full \textit{RXTE} archive is available, this dataset provides the highly stable, long-baseline folded template required to average over short-term flaring and isolate the persistent geometric modulation. Given the lack of formal error bars in the long-term folded template, we assigned a conservative constant uncertainty of 5\% the count rate to each phase bin. This uniform weighting ensures that the fit is driven by the global morphology of the modulation rather than by bin-to-bin fluctuations in the folded lightcurve.

\begin{figure*}[t]
\centering
\includegraphics[width=\textwidth]{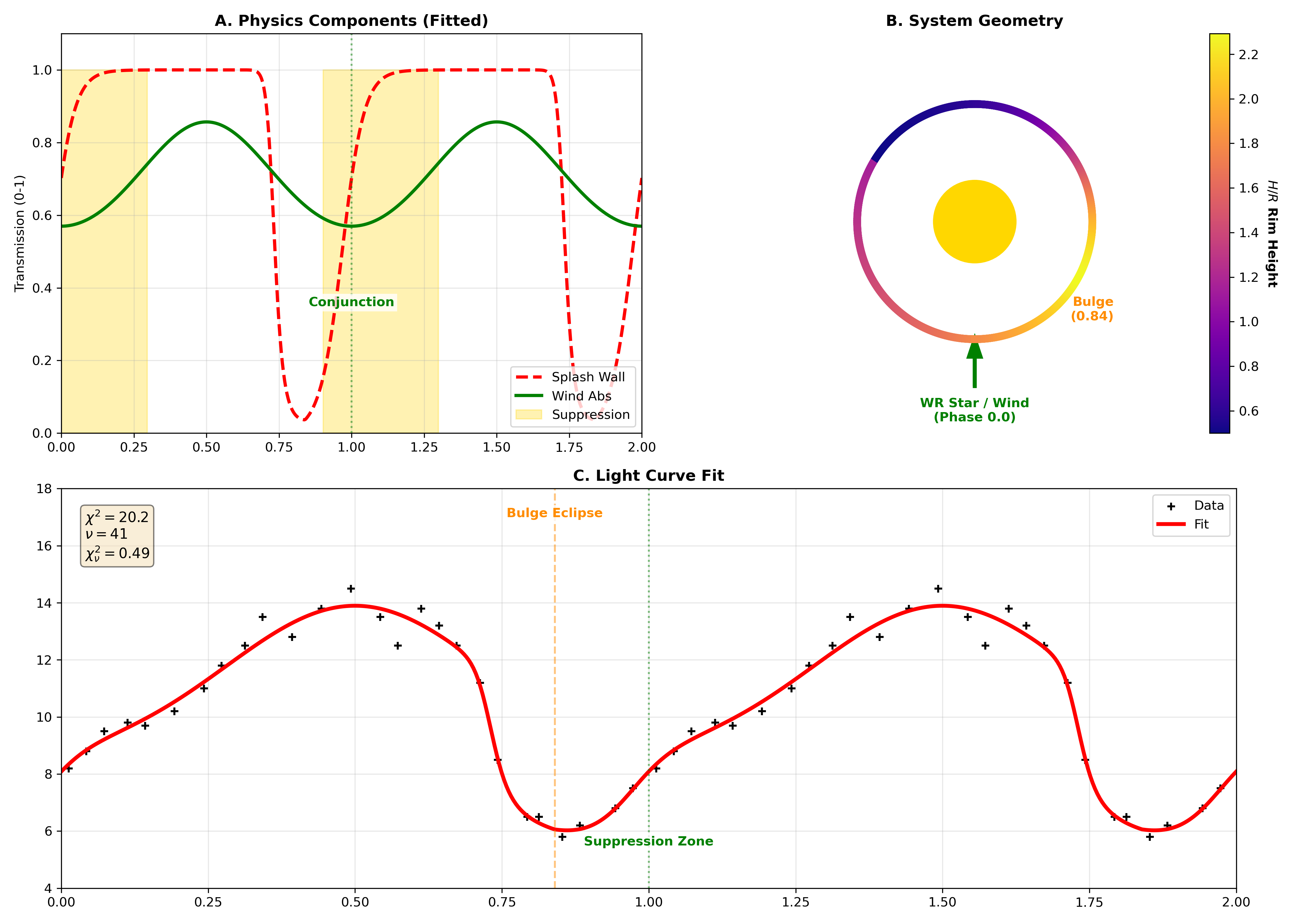} 
\caption{\textbf{Physical modeling of the Cygnus X-3 orbital modulation.} 
\textbf{(A)} Decomposition of the normalized transmission components. The dashed red line shows the geometric visibility driven by the ``Turbulent Wall'' (Stream Impact Bulge), which blocks the extended X-ray source. The solid green line shows the modulation from the Wolf--Rayet wind absorption, centered at superior conjunction ($\phi=0.0$, vertical dotted line). The yellow shaded region highlights the ``Suppression Zone,'' where the geometric egress overlaps with the wind wake. \textbf{(B)} Top-down view of the accretion disk geometry in the physical frame ($\phi=0.0$ is the Wolf--Rayet closest approach). The color scale indicates the variable rim height $H(\phi)/R$. The asymmetric bulge is localized to $\phi_{wall} \approx 0.84$, consistent with the ballistic impact of the accretion stream. \textbf{(C)} The optimized model fit (red solid line) to the folded \textit{RXTE}/ASM lightcurve (black crosses). The data have been shifted by $\Delta\phi = -0.11$ to align the observed minimum with the physical frame. The model successfully reproduces the asymmetric ``fast-fall, slow-rise'' morphology through the combined effects of the sharp geometric ingress and the broad wind absorption wake. For reproducibility purposes the Python code used for this modeling is given in Appendix \ref{app:code}}.
\label{fig:model_fit}
\end{figure*}

We model the rim height $H(\phi)$ as a localized perturbation, visualized in the physical frame in Figure \ref{fig:model_fit}B, defined by:
\begin{equation}
H(\phi)=H_{base}+H_{amp}\cdot S(\phi)
\end{equation}
where the shape function $S(\phi)$ describes the hydrodynamic profile of the bulge. 

In this formalism, we normalize the outer disk radius to unity, such that the vertical parameters $H_{base}$ and $H_{amp}$ represent the dimensionless aspect ratios ($H/R$) rather then absolute physical lengths. To reproduce the ``fast ingress / slow egress'' morphology observed in the data, we utilize a composite profile characterized by a Gaussian shock front and an exponential downstream wake:
\begin{equation}
    S(\phi) = 
    \begin{cases} 
    \exp \left( -0.5 \left( \frac{\phi - \phi_{wall}}{\sigma_{shock}} \right)^2 \right) & \text{if } \phi < \phi_{wall} \\
    \exp \left( - \frac{|\phi - \phi_{wall}|}{\lambda_{decay}} \right) & \text{if } \phi \ge \phi_{wall}
    \end{cases}
\end{equation}
where $\phi_{wall}$ is the orbital phase of the impact peak. The transmission $T_{geo}$ is derived using a sigmoid function to approximate the fractional visibility of the extended source:
\begin{equation}
    T_{geo}(\phi) = \left[ 1 + \exp\left( -k \cdot (\theta_{rim}(\phi) - i) \right) \right]^{-1}
\end{equation}
where $\theta_{rim}(\phi) = \arctan(H(\phi)/R_{disk})$ and $k$ is the 'hardness' parameter, which governs the steepness of the flux transition. Physically, $k$ acts as a proxy for the vertical angular extent of the X-ray source perpendicular to the occulting rim. In this formalism, a high value ($k \gg 10$) approximates a point source eclipsed by a sharp edge, resulting in a step-function ingress. Conversely, a low value ($k \sim 1$) describes a vertically extended scattering photosphere  that is gradually covered by the rising Wall. The angular size of the source is approximately related to the hardness by $\theta_{source} \approx 4/k$. Thus, $k$ allows the lightcurve morphology to discriminate between a compact black hole corona and the large-scale scattering funnel predicted by polarimetry.

We define the observed flux at orbital phase $\phi$ as:
\begin{equation}
    L(\phi) = \left[ F_{min} + (F_{max} - F_{min}) \cdot T_{geo}(\phi) \right] \cdot T_{wind}(\phi)
\end{equation}
where $F_{min}$ represents the unocculted DC scattering halo flux, and $F_{max}$ corresponds to the peak intrinsic flux of the core. 

The Wind Transmission, $T_{wind}(\phi)$, is modeled as a phase-dependent optical depth variation centered at the Wolf--Rayet superior conjunction:
\begin{equation}
    T_{wind}(\phi) = \exp \left( - \left[ \tau_{avg} + \tau_{amp} \cos(2\pi \phi) \right] \right)
\end{equation}
where $\tau_{avg}$ represents the mean optical depth of the wind along the line of sight, and $\tau_{amp}$ defines the amplitude of the modulation due to the varying path length through the Wolf--Rayet atmosphere. This simple cosine form captures the broad, smooth absorption component that governs the lightcurve egress, distinct from the sharp geometric occultation of the ingress. We did not attempt to model any density variations with radius in the WR wind, due to uncertainties in its ionization state and velocity profile.

\subsection{Results}

The results of the non-linear least-squares minimization fit are presented in Table \ref{tab:model_params_final} and Figure \ref{fig:model_fit}. The contribution of the two normalized transmission components are shown in Figure \ref{fig:model_fit}A. To derive parameter uncertainties, we performed a residual bootstrap analysis, generating 100 synthetic datasets by resampling the residuals of the best-fit model with replacement. Each synthetic lightcurve was re-fitted with all parameters left free. The resulting $1\sigma$ bootstrap uncertainties are also given in Table \ref{tab:model_params_final}

A critical free parameter in our analysis was the global phase shift, $\Delta\phi$, which aligns the observed X-ray minimum with the physical coordinate frame defined by the Wolf--Rayet superior conjunction. The optimization yields a robust phase shift of $\Delta\phi = -0.11 \pm 0.01$, confirming that the observed X-ray minimum lags the physical conjunction. The ``Turbulent Wall'' is localized to a physical orbital phase of $\phi_{wall} = 0.84 \pm 0.02$. The deviation of this value from the canonical circularization phase ($\phi_{circ} \approx 0.89$) represents a physical `early impact' of $\Delta\phi \approx 0.05$. This confirms that the accretion disk is radially extended, intercepting the stream near the tidal truncation radius before it completes its ballistic trajectory. These parameters establish a stratified modulation where a sharp, Gaussian-like ingress ($\sigma_{shock} \approx 0.14$) represents the impact of the accretion stream, followed by a more gradual exponential decay ($\lambda_{decay} \approx 0.49$) as the turbulent wake dissipates around the disk rim.

The hardness parameter $k=0.74$ implies a vertical source height of $\theta_{source} \approx 5.4^{\circ}$. Because of the gradual slope of the occulting Wall, it takes the system approximately $\Delta\phi \approx 20^{\circ}$ of orbital phase to eclipse this $5.4^{\circ}$ structure. This physical size is fully consistent with the \textit{IXPE} polarization constraints, which define a maximum funnel width of $\sim 30^{\circ}$. 

The fitted parameters imply a peak vertical extent of the ``Turbulent Wall'' of $(H/R)_{\rm peak} \simeq H_{\rm base}+H_{\rm amp} \approx 2$, indicating that the occulting structure is geometrically thick and extends well above the disk midplane. This vertical extent does not arise from hydrostatic disk inflation, but from the localized kinetic energy deposition of a super-Eddington RLOF stream impacting the outer disk rim.

The fit also identifies an intrinsic flux floor ($F_{min} \approx 9.4$ counts s$^{-1}$) that remains unocculted even during the deepest geometric minimum. This is likely from both Thompson scattering in the WR wind and the extended ``Scattering Mirror'' higher up in the funnel, responsible for the \textit{IXPE} polarized signal. This configuration acts as a ``coronagraph,'' which results in increased polarization degree during minimum observed by \textit{IXPE}. Disentangling the contribution of the ``Scattering Mirror'' from that of the wind scattering is not possible within our model because the two components are degenerate in the flux data. 

This model confirms that the ``Turbulent Wall'' acts as a natural coronagraph, selectively blocking the compact $\approx 16.2$ counts s$^{-1}$ core while leaving the extended scattering mirror and ambient wind halo visible to the observer.

A central result of our modeling is the decoupling of the observational X-ray ephemeris from the physical binary clock ($\Delta\phi$), and it is important to distinguish between two phase definitions. The X-ray Phase ($\phi_{X}$) is the standard observational definition, where $\phi_{X}=0.0$ corresponds to the X-ray minimum \citep{vanderKlis1989,Choudhury2004}. The Binary Phase ($\phi_{bin}$) is the physical dynamical frame, where $\phi_{bin}=0.0$ corresponds to the superior conjunction of the Wolf--Rayet donor (see Figure \ref{fig:model_fit}).

While our model successfully captures the global geometry of the system, we note that several parameters (such as the base rim height and source hardness) exhibit large formal uncertainties. These uncertainties are primarily systematic, arising from geometric degeneracies and covariance between the model parameters (e.g., viewing angle versus vertical structure height), rather than limitations in the counting statistics of the current dataset.

\begin{table}[h]
\centering
\caption{Optimized Physical Model Parameters (All Fitted)}
\label{tab:model_params_final}
\begin{tabular}{lcc}
\hline\hline
\textbf{Parameter} & \textbf{Symbol} & \textbf{Value} \\
\hline
\multicolumn{3}{c}{\textit{System Geometry}} \\
\hline
Orbital Inclination & $i$ & $28.0^{\circ}$ (Fixed) \\
Phase offset ($\phi_{bin}$ $\rightarrow$ $\phi_X$ )  & $\Delta\phi$ & $-0.11 \pm 0.01$ \\
\hline
\multicolumn{3}{c}{\textit{Turbulent Wall (Stream Impact Bulge)}} \\
\hline
Wall Peak Phase & $\phi_{wall}$ & $0.84 \pm 0.02$ \\
Base Rim Height & $H_{base}/R_{disk}$ & $0.50 \pm 0.47$ \\
Bulge Amplitude & $H_{amp}/R_{disk}$ & $1.79 \pm 0.79$ \\
Shock Width (Ingress) & $\sigma_{shock}$ & $0.14 \pm 0.04$ \\
Decay Scale (Egress) & $\lambda_{decay}$ & $0.49 \pm 0.16$ \\
Source Hardness & $k$ & $0.74 \pm 0.52$ \\
\hline
\multicolumn{3}{c}{\textit{Wolf--Rayet Wind}} \\
\hline
Average Optical Depth & $\tau_{wind, avg}$ & $0.36 \pm 0.03$ \\
Wind Modulation Amplitude & $\tau_{wind, amp}$ & $0.20 \pm 0.02$ \\
\hline
\multicolumn{3}{c}{\textit{Flux Levels (ASM Counts s$^{-1}$)}} \\
\hline
Intrinsic Minimum (DC Halo) & $F_{min}$ & $9.42 \pm 0.80$ \\
Intrinsic Maximum (Core) & $F_{max}$ & $16.22 \pm 0.45$ \\
\hline
\multicolumn{3}{c}{\textit{Fit Statistics}} \\
\hline
Total Chi-Squared & $\chi^2$ & $20.2$ \\
Degrees of Freedom & $\nu$ & $41$ \\
Reduced Chi-Squared & $\chi^2_\nu$ & $0.49$ \\
\hline\hline
\end{tabular}
\tablecomments{Values derived from 100 bootstrap iterations. The reduced chi-squared of 0.49 indicates a satisfactory fit to the global morphology, given the constant 5\% error assumption applied to the ASM data.}
\end{table}

\section{Kinematics and the Binary Mass}
\label{sec:kinematics}

X-ray spectroscopy provides constraints on the location of the absorbing and emitting plasma, which in turn provides a critical test of the ``Hybrid RLOF" geometry. High-resolution mapping confirms that the majority of the lower-energy X-ray lines (e.g., from Si, S, Ar, and Ca) originate deep within the dense Wolf--Rayet wind, in close proximity to the donor star \citep{Kallman2019, Vilhu2025}. In contrast, the highly ionized Iron K-shell emission (Fe {\sc xxv} and Fe {\sc xxvi}) exhibits distinct orbital variability associated with the compact object \citep{Miura2025}, proving it arises from a physically separate, extreme-ionization region.

The observed width of the Fe\,{\sc xxvi} emission line (FWHM $\approx 800$\,km~s$^{-1}$; \citealt{Miura2025}) is significantly narrower than the terminal velocity of the Wolf--Rayet wind ($v_{\infty} \approx 1700$\,km~s$^{-1}$). This provides an important constraint on their origin. Emission from the inner accretion flow ($R < 100 R_g$) would be Doppler-broadened beyond recognition by extreme Keplerian velocities ($v \gtrsim 0.1c$). Conversely, an origin in an inner, fast-outflowing black hole wind would induce terminal absorption velocities of several thousand km~s$^{-1}$. Instead, \textit{XRISM} observes a highly phase-dependent outflow with absorption velocities of only $-500$ to $-600$\,km~s$^{-1}$. Therefore, the relatively narrow $\sim 800$\,km~s$^{-1}$ emission profile and the moderate outflow velocity require an origin at much larger radii. 

Crucially, the lack of a double-peaked Keplerian profile confirms that the emission is not distributed symmetrically around the disk, but is localized to a specific site. The obvious location is the turbulent stream-disk impact shock at the outer rim of the accretion disk \citep{Armitage1998}. This is confirmed by the \citet{Miura2025} results. They find that the radial velocity of the Fe {\sc xxvi} lines crosses zero at $\phi_X = -0.02\, (+0.05, -0.04)$. In our model, this corresponds to the ``Turbulent Wall'' crossing the line of sight, where its projected rotational velocity naturally vanishes. The fact that the line velocity nulls at the Wall's location rather than at the model-inferred dynamical conjunction ($\phi_X \approx 0.11$) confirms that the emitter is tied to the non-axisymmetric accretion stream. The velocity width directly measures the violent turbulence in this shock layer. The survival of strong iron-line emission at $\phi_X = 0.0$, when the inner wall is hidden, requires that the line-emitting atmosphere rises above the occulting Wall in a wind/corona/photosphere. This framework naturally resolves the apparent tension between the extreme densities required for continuum occultation and the moderate densities inferred from X-ray spectroscopy by separating the dynamical anchor of the emission (the dense Wall) from the extended, lower-density atmosphere that dominates the observed Fe\,{\sc xxvi} Ly$\alpha$ flux.

As the system moves past $\phi_X \approx 0.125$, the Wall rotates to the far side of the orbit, such that line emission from the Bulge must traverse the dense central regions of the Wolf--Rayet wind before reaching the observer. This prolonged ``wind eclipse of the Bulge'' reproduces the wide suppression of the Fe {\sc xxvi} Ly$\alpha$ line, the ``Iron Dip'' \citep{Kitamoto1994}, and the high-velocity absorption observed by \textit{XRISM} \citep{XRISM2024}.

With the emission securely tied to the stream-impact site, we can estimate the black hole mass. For an inclination of $i \approx 28^{\circ}$ \citep{Veledina2024a}, the radial velocity amplitude of the Fe\,{\sc xxvi} lines in Cygnus X-3 is $K_{obs} = 430^{+150}_{-140}$ km s$^{-1}$ \citep{Miura2025}. The observed velocity amplitude is the sum of the black hole's projected orbital semi-amplitude ($K_{BH} = V_{orb} \sin i$) and the projected rotational velocity of the Wall ($V_{wall} \sin i$):
\begin{equation}
K_{obs} \approx K_{BH} + V_{wall} \sin i .
\end{equation}

For a canonical low-mass black hole ($M_X \approx 5 M_\odot$) paired with a $20 M_\odot$ donor ($q \approx 4$), the orbital speed is so high that $K_{BH}$ alone would exceed 500 km s$^{-1}$, violating the observed nominal kinematics. In fact, matching the observation in this scenario would require the unphysical condition of retrograde disk rotation. In contrast, a massive black hole of $M_X \approx 15 M_\odot$ ($q \approx 1.3$) balances the budget. This mass yields a projected orbital amplitude of $K_{BH} \approx 240$ km s$^{-1}$. The difference between this and the observed nominal $430$ km s$^{-1}$ is supplied by the Wall's rotational velocity ($V_{wall} \sin i \approx 190$ km s$^{-1}$), which corresponds to a physical flow speed in the disk plane of $V_{wall} \approx 400$ km s$^{-1}$. For a $15 M_\odot$ black hole, the local Keplerian velocity at the tidal truncation radius ($R_{disk} \approx 1.4 R_\odot$) is $V_K \approx 1430$ km s$^{-1}$. The derived Splash Wall velocity ($\approx 400$ km s$^{-1}$) is therefore highly sub-Keplerian ($\approx 28\%$ of $V_K$). This provides decisive kinematic evidence that the line-emitting material is not in a stable free-fall orbit. Instead, it represents the decelerated, post-shock stream at the circularization radius of a wind-fed disk, where specific angular momentum is naturally low.

Crucially, the inferred mass of the Wolf--Rayet star is consistent with historical interpretations of the infrared radial velocity curves. \citet{Hanson2000} noted that the radial velocity zero-crossing of the He\,{\sc i} lines lags the X-ray minimum, and that the velocity amplitude measured from these infrared features is systematically underestimated. We caution that the mass function derived by \citet{Hanson2000} is likely unreliable. Their analysis relied on the He\,{\sc i} 2.058\,$\mu$m line during an outburst, where complex P-Cygni variability and wind ionization changes can severely distort the apparent radial velocity of the absorption component. As discussed by \citet{Zdziarski2013} and further supported by detailed spectral modeling of the wind structure \citep{Koljonen2017}, irradiation and wind-shielding effects shift the effective line-formation region toward the $L_1$ point, significantly reducing the observed orbital velocity swing. Accounting for this infrared ``RV offset,'' together with the Splash Wall geometry identified here, yields system parameters that are fully consistent with a $\sim20\,M_{\odot}$ Wolf--Rayet donor and a $\sim15\,M_{\odot}$ black hole.

\section{Discussion}
\label{sec:discussion}

\subsection{Resolving the Geometric Crisis}
\label{sec:Crisis}
The ``Geometric Crisis'' effectively offers a choice between a wind-formed Accretion Wake or an RLOF-formed Stream Bulge. The ``Hybrid'' RLOF plus disk-bulge geometry reproduces the depth, asymmetry, and non-zero flux floor of the orbital X-ray modulation without invoking the extreme wind opacities, accretion wake and shocks required by the wind accretion model. The key insight is that in this ``Hybrid'' RLOF model, the observed modulation is produced by two physically distinct processes operating at different orbital phases (Figure \ref{fig:model_fit}). A vertically extended, azimuthally localized ``Turbulent Wall'' at the outer disk rim provides the geometric shutter that controls the depth of the minimum ($\phi_X \approx 0.0$). Meanwhile, phase-dependent attenuation in the dense Wolf--Rayet wind governs the slow recovery, producing a ``suppression zone'' where the bulge has rotated out of the line of sight, but the wind takes over as a scattering medium with moderate optical depth ($\tau \sim 0.4$).

In this scenario, the opaque shutter (the ``Turbulent Wall'') selectively occults the unpolarized central emission, increasing the fractional contribution of the polarized reflection from the surrounding funnel walls. The \textit{IXPE} detection of a polarization dip at the 6.4 keV Iron line provides the definitive signature of this reflection geometry \citep{Veledina2024a}. 

This model also naturally explains the behavior during High/Soft states, where the PD is observed to drop significantly. In the wind model, a higher accretion rate would increase the wind density, potentially increasing scattering and polarization. Conversely, in the ``Wall'' model, the increased mass transfer enhances the radiative heating within the scattering funnel, flooding the system with unpolarized soft X-ray thermal emission. This spectral dilution reduces the net observed polarization without requiring a change in the scattering geometry itself.

The geometric invariance of the lightcurve---its tendency to remain stable despite dramatic transitions between High/Soft and Low/Hard states---is a natural consequence of the ``Hybrid'' RLOF model. Because the position of the accretion stream impact is fixed by the binary's orbital parameters (specifically the $L_1$ trajectory), the resulting bulge remains a permanent geometric fixture \citep{Lubow1975, Lubow1976}. Unlike wind-absorption models, where the optical depth $\tau$ would fluctuate wildly with changes in mass-loss rate and ionization, the super-critical bulge is inherently opaque. Thus, the modulation is governed by the shutter's projected area (a geometric constant) rather than the variable density of the gas, ensuring the lightcurve template remains consistent even as the system's total luminosity varies by orders of magnitude.

For comparison, it is instructive to consider whether BHL wind accretion can satisfy the same geometric and energetic constraints. In a 4.8~hr orbit, the accretion luminosity depends sensitively on the wind velocity at the compact object, scaling approximately as $L_{\rm BHL} \propto \dot{M}_{\rm w} v_{\rm w}^{-4}$ \citep[e.g.,][]{Pringle1974}. Hydrodynamic simulations show that a WN7 Wolf–Rayet star approaching—but not necessarily filling—its Roche lobe naturally reshapes a quasi-spherical WR wind into a slow, dense, equatorially focused flow that dramatically enhances accretion \citep[e.g.,][]{Friend1982, Hadrava2012, ElMellah2019}. Assuming a standard mass-loss rate of $\dot{M}_{\rm wind} \approx 3 \times 10^{-5}\,M_{\odot}\,{\rm yr}^{-1}$ for a WN7 Wolf--Rayet star \citep{Hamann2006}, simulations of wind-focused Roche-lobe overflow demonstrate that up to $\sim 20\%$ of the wind can be gravitationally focused into the accretion stream \citep{ElMellah2019}, yielding an accretion rate of $\approx 6\times 10^{-6}\,M_{\odot}\,{\rm yr}^{-1}$. 

For a $15\,M_\odot$ black hole undergoing helium accretion, the Eddington luminosity is $L_{\text{Edd}} \approx 4 \times 10^{39}$,erg,s$^{-1}$. While a standard thin disk could achieve this with an accretion rate of roughly $\sim 7 \times 10^{-7}\,M_\odot\,\text{yr}^{-1}$ (about 7\% of the wind mass-loss rate), the high transfer rate provided by the focused wind drives the inner flow into a ``slim'', advection-dominated regime. In this puffed-up state, the radiative efficiency decreases rapidly (as detailed in Appendix~\ref{sec:Appendix_Orbital}). Therefore, rather than over-producing X-rays, this higher capture fraction ($\sim 20\%$) and resulting mass transfer rate are strictly necessitated to power the observed luminosity. We adopt an efficiency parameter of $\beta = 0.01$ \citep[e.g.,][]{Abramowicz1988}.

Achieving the implied super-Eddington luminosities of $L_X \ge 10^{39}\,\mathrm{erg\,s^{-1}}$ therefore requires either unusually slow winds or extreme Wolf--Rayet mass-loss rates. However, detailed wind-accretion models show that such conditions inevitably produce order-unity optical depths, strong phase-dependent column density variations, and extended accretion wakes \citep{Okazaki2014,Vilhu2021}, in tension with the observed ``grey'' modulation and the lack of significant orbital $N_{\rm H}$ variability \citep{Zdziarski2012}. Conversely, adopting faster, highly ionized winds consistent with high-resolution X-ray spectroscopy yields accretion rates insufficient to power the source \citep{Vilhu2021}. 

In the ``Hybrid'' RLOF geometry the Wolf--Rayet wind is not required to supply the full optical depth responsible for the X-ray minimum. The dominant occultation is provided by the dense ``Turbulent Wall'' at the disk rim, while the wind contributes a secondary, extended scattering component. As a result, the maximum effective optical depth required of the wind is reduced by approximately a factor of two relative to pure wind-absorption models, alleviating the ionization and column density constraints that challenge the BHL scenarios. This intrinsic tension highlights why a purely wind-fed geometry struggles to reproduce the observed combination of luminosity, ``grey'' modulation, and polarization, whereas the ``Hybrid'' RLOF scenario intrinsically satisfies all three.
\label{sec:Wall}
\subsection{Physical Origin and Vertical Extent of the Turbulent Wall}

The ``Turbulent Wall'' $(H/R)_{peak}\sim 2$ may seem extreme compared to standard, radiation-supported `slim disk' models. Indeed, puffing up the outer disk via internal radiation pressure alone would require implausibly large accretion rates ($\dot{M} \approx 10^5 \dot{M}_{Edd}$) \citep{Liu2025}. However, the occulting wall geometry is not a product of internal radiation pressure. Rather, it is a dynamic feature formed by the direct kinetic impact of the RLOF stream colliding with the outer rim of the accretion disk \citep{Armitage1998}. This stream-disk impact creates a localized, shock-heated bulge---a well-documented phenomenon responsible for the orbital X-ray dips observed in LMXBs such as EXO 0748-676 \citep{Parmar1986}. 

In standard LMXB ``Dippers'' e.g., \citet{DiazTrigo2006}, the orbital modulation arises when the line of sight ($i \sim 60^{\circ}-80^{\circ}$) intersects the raised rim of the accretion disk. In Cygnus X-3, the interaction is more extreme: the stream impact is so violent that the resulting bulge extends to high latitudes. We refer to this as a ``Polar Dipper'', where we observe dipping behavior—periodic occultation of the central engine—even when viewing the system at $i \approx 28^{\circ}$.

The high density required for the Wall ($n_e \sim 10^{15}$\,cm$^{-3}$) is physically consistent with these mass flow rates. Our model fit indicates a ``Turbulent Wall'' height (disk plus bulge) of $H \sim 2 R_{disk}$ and an azimuthal shock width of $\approx 0.14$ in phase. Assuming the stream penetrates with a radial thickness of $\Delta R \sim 0.1R_{\text{disk}}$ (a conservative upper limit based on the typical supersonic stream width of $\lesssim 0.05 R_{\text{disk}}$), implies an impact area $A \approx H \times \Delta R \sim 2 \times 10^{21}$\,cm$^2$ (for $R_{disk} \approx 10^{11}$\,cm). For a local stream transfer rate of $\dot{M} \approx 1.5 \times 10^{-5}\,M_\odot$\,yr$^{-1}$ impacting at a velocity $v \approx 400$\,km\,s$^{-1}$ (consistent with the kinematic line widths), continuity implies $\rho \approx \dot{M}/(vA) \sim 1.2 \times 10^{-8}$\,g\,cm$^{-3}$, corresponding to $n_e \approx 4 \times 10^{15}$\,cm$^{-3}$ for a Helium dominated flow. This confirms that the stream impact site naturally generates the high densities required for continuum occultation.

The observed increase in iron line equivalent width during X-ray minimum \citep{Serlemitsos1975,XRISM2024} is a natural consequence of this geometry. The ``Turbulent Wall'' effectively acts as a coronagraph, occulting the compact continuum core ($R_{\rm core} \sim 10^8$\,cm) at $\phi_X = 0.0$ while leaving the large-scale, line-emitting halo visible. This phenomenology mirrors the ``eclipse behavior'' observed in classical HMXBs (e.g., Cen~X-3, Vela~X-1), where the equivalent width of the fluorescent iron line rises sharply as the neutron star passes behind the donor, suppressing the direct continuum while the extended wind nebula remains illuminated \citep{Aftab2019}. 

This geometric separation is supported by the phase-resolved behavior of the Fe\,{\sc xxvi} Ly$\alpha$ line observed with \textit{XRISM} \citep{XRISM2024}. The data reveal a broad suppression of the line flux from $\phi_X \approx 0.125$ to $0.50$. In our ``Hybrid'' model, this interval corresponds to orbital phases where the ``Turbulent Wall'' ($\phi_{\rm phys} \approx 0.84$) lies on the far side of the binary relative to the observer, such that the line of sight to the emitting Bulge must traverse the densest regions of the Wolf--Rayet wind (centered at $\phi_{\rm phys} \approx 0.0$). Conversely, the line flux reaches its observed maximum at the opposite phase ($\phi_X \approx 0.65$--$0.85$), when the Bulge rotates to the near side of the system, providing a direct and unobscured view of the emission region. In this framework, the long-recognized ``iron dip'' \citep{Kitamoto1994} is physically identified as a prolonged wind eclipse of the accretion-stream bulge.

The analogy with other RLOF systems can be extended to CVs, where a similar disk/stream interaction occurs. The ``Hot Spot'' in a CV produces emission lines with velocity vectors deviating from the white dwarf's motion (the classic ``S-wave''; \citealt{Warner1995}, Marsh1988), the ``Turbulent Wall'' in Cygnus X-3 introduces a non-Keplerian velocity component to the Iron line profiles. The \textit{XRISM} detection of a large velocity amplitude ($K_{obs} \approx 430$ km s$^{-1}$) is therefore not a direct measurement of the black hole's orbital speed, but a tracer of the turbulent, high-velocity impact region. 

The detailed microstructure of the ``Turbulent Wall'' warrants further investigation. The interaction between the supersonic accretion stream and the super-critical disk rim involves complex non-linear fluid dynamics, including radiative shocks and Kelvin-Helmholtz instabilities, which are beyond the scope of this work. We identify full 3D radiation-hydrodynamic simulations as the critical next step to confirm the vertical stability of the impact bulge and the dynamics of the matter ejected from the system.

\label{sec:RLOF}
\subsection{Does the Wolf--Rayet Donor Fill its Roche Lobe?}

Whether the Wolf--Rayet (WR) donor in Cygnus~X-3 fills its Roche-lobe depends critically on its spectral classification and the resulting definition of its stellar radius. Historically, the low mass function derived from infrared wind lines ($f(M) \approx 0.027 M_{\odot}$; \citealt{Hanson2000}) implied a low-mass companion. However, the low inclination constrained by \textit{IXPE} ($i \approx 28^{\circ}$) implies a much higher de-projected orbital velocity for the WR star. Assuming the compact object is a black hole, this geometry requires the companion to be a massive Wolf--Rayet star ($M_{WR} \gtrsim 15\text{--}20 M_{\odot}$).

This mass requirement strongly favors the WN~5--7 spectral classification \citep{vanKerkwijk1996} over the lighter WN~4--6 type \citep{Koljonen2017}. While WN~4 stars are typically low-mass ($\sim 5 M_{\odot}$), Galactic WN~7 stars are intrinsically massive, with typical masses of $M_{WR} \approx 20$--$50 M_{\odot}$ \citep{Crowther2007}. We attribute the observed high-ionization lines (e.g., He II and N V), which formally suggest an earlier WN subtype, to X-ray photoionization of the stellar wind by the compact object. Phase-resolved IR spectroscopy demonstrates that these lines originate in the X-ray–modified wind rather than in the stellar photosphere, and thus their ionization balance reflects external irradiation rather than the intrinsic effective temperature of the donor \citep{vanKerkwijk1996, Koljonen2017}. 

For a WN~7 donor, the classical hydrostatic core radius ($R_{\rm core}\lesssim1\,R_\odot$; \citealt{Langer1989}) is not the relevant scale for mass transfer. Instead, the effective photosphere forms in the accelerating optically thick wind. Modern non-LTE atmosphere models place this photospheric radius substantially further out, at $R_{\rm phot}\sim1.5$--$3\,R_\odot$ (e.g., \citealt{Hamann2006, Grafener2012}).

For an orbital period of 4.8~hr and a total system mass of $\sim20$--$25\,M_\odot$ (assuming $M_{\rm X} \sim 5 M_{\odot}$), the binary separation is $a\simeq3$--$3.5\,R_\odot$. Using Eggleton’s approximation with a mass ratio $q=M_{\rm WR}/M_{\rm X}\simeq3$--$4$, the Roche-lobe radius of the WR donor is $R_{\rm RL}\simeq1.4$--$1.7\,R_\odot$. Thus, while the hydrostatic core underfills the Roche-lobe, the effective wind photosphere of a WN~7 donor grazes or overfills it. In this full RLOF regime, we expect the compact object to capture approximately 50\% of the mass loss (the hemisphere facing $L_1$), yielding a transfer rate of $\approx 1.5 \times 10^{-5} M_{\odot}$ yr$^{-1}$.

This conclusion is robust even for the higher black hole mass ($M_{\rm X} \sim 15 M_{\odot}$) suggested by our kinematic analysis of the velocity budget (Section \ref{sec:kinematics}). Increasing the black hole mass reduces the mass ratio $q$, which decreases the fractional Roche-lobe size ($R_{RL}/a$), but simultaneously increases the binary separation $a$. These effects largely offset each other. For $M_{\rm X} = 15\,M_\odot$, the Roche-lobe radius remains comparable to the WN~7 photospheric radius. With a mass ratio of $q \approx 1.3$, the system remains comfortably below the critical threshold for dynamical instability ($q_{\rm crit} \sim 3.5$) for donors with radiative envelopes, ensuring that the Roche-lobe overflow proceeds stably on the nuclear timescale rather than running away \citep{VanDenHeuvel2017}.

\subsection{Orbit Evolution}
\label{sec:Orbit}
Historically, the observed orbital expansion of Cygnus~X-3
($\dot{P}/P \approx 1.0 \times 10^{-6}\,{\rm yr}^{-1}$; \citealt{Singh2002,Zdziarski2018})
has been used to argue against Roche-lobe overflow, since conservative mass transfer from a
heavier donor would shrink the orbit \citep{Savonije1978,VanDenHeuvel2017}. In the highly
super-critical regime considered here, however, the evolution is strictly non-conservative:
extreme mass loss from the Wolf--Rayet donor weakens the gravitational potential and
naturally drives orbital expansion (Appendix~\ref{sec:Appendix_Orbital}). For Keplerian
orbits $a \sim j^2/(GM)$, so orbital separation responds linearly to fractional mass loss,
whereas comparable changes driven by angular-momentum loss require large-scale
redistribution prior to escape.

The observed expansion, together with the super-Eddington luminosity and highly ionized
outflows, indicates that most non-accreted material is expelled from the inner accretion
flow. We show in Appendix~\ref{sec:Appendix_Orbital} that the measured $\dot{P}$ is reproduced by the same super-Eddington transfer
rates ($\approx 2 \times 10^{-5}\,M_{\odot}\,{\rm yr}^{-1}$) required by the Splash Wall
geometry (Section~\ref{sec:Crisis}), directly linking the orbital evolution to the inner-disk
accretion physics.

This interpretation is supported by global radiation-MHD simulations of super-Eddington
accretion, which show that radiation-dominated disks become geometrically thick
($H/R \sim 1$) and eject the majority of the supplied mass via powerful winds launched from
$\sim10$--$100\,R_g$ \citep{Ohsuga2011,Sadowski2014}. In such flows, angular momentum is
removed locally by disk winds over a broad range of radii, preventing efficient storage at
large disk radii or recycling back into the binary. Although tidal torques from the
Wolf--Rayet donor can return angular momentum to the orbit \citep{Papaloizou1977}, these act
only on material that remains bound and reaches the truncation radius, whereas in Cygnus~X-3
the dominant mass loss occurs well inside this region.

Because these outflows originate deep in the potential well, they carry modest specific
angular momentum relative to the binary orbit ($\alpha \sim 1$), providing the low-$j$
mass--loss-channel required to sustain $\dot P>0$. Thus, the super-Eddington luminosity,
phase-resolved iron-line diagnostics, and secular orbital expansion collectively indicate
that most non-accreted mass is expelled from the inner accretion flow in disk-driven winds
and jet-associated outflows.

Although secular orbital expansion can in principle arise from Wolf--Rayet wind mass loss
in classical BHL accretion, such models face severe quantitative and
phenomenological challenges in Cygnus~X-3. Reproducing the observed X-ray luminosity via
wind capture alone requires unrealistically large donor mass-loss rates and capture
efficiencies for the inferred wind velocities. Moreover, the observed wind velocities imply capture fractions well below those required to sustain the measured luminosity. We refer here to the local wind velocity at the orbital separation, not the terminal velocity ($v_\infty$). Given the highly compact nature of the Cygnus X-3 orbit, the Wolf-Rayet wind is still accelerating and has not yet reached terminal speed when it encounters the compact object. This low local relative velocity pushes the accretion mechanism beyond classical Bondi-Hoyle-Lyttleton capture. Instead, it facilitates a transition into a 'wind-RLOF' or hybrid Roche-lobe overflow regime, where the gravitational potential strongly focuses the slow wind, making the required high accretion rates naturally attainable. More fundamentally, pure wind-fed scenarios do not naturally produce the geometrically thick inner disk, phase-resolved iron-line structure, or powerful radiation-driven outflows that are directly observed. Even allowing for supercritical accretion in a wind-fed geometry, the required inner-disk winds and re-ejection effectively converge toward a Roche-lobe overflow–like configuration, with mass supplied to and expelled from the inner accretion flow.

In contrast, Roche-lobe overflow directly feeds the compact object, naturally generating a
supercritical disk and disk-driven winds that simultaneously account for the luminosity,
spectral diagnostics, inferred inner-disk geometry, and secular orbital expansion. As shown
in Appendix~\ref{sec:Appendix_Orbital}, Cygnus~X-3 robustly occupies the mass--loss-dominated
expansion regime even under conservative uncertainties, demonstrating that the observed
$\dot P>0$ follows directly from extreme non-conservative mass transfer. We therefore
conclude that highly supercritical Roche-lobe overflow provides not merely a viable
alternative, but the most self-consistent framework for the observed properties of
Cygnus~X-3.

\subsection{Multi-Wavelength Consistency}

The ``Hybrid'' geometry predicts the energy dependence of the orbital modulation observed above 30 keV. At soft X-ray energies ($\sim$1--10~keV), the modulation arises from the combination of two physically distinct components: geometric occultation by the ``Turbulent Wall'' (sharp ingress) and Thompson scattering and iron K photoelectric absorption in the dense Wolf--Rayet wind (broad egress). 

As photon energy increases, the Iron K absorption becomes irrelevant above $\sim 30$~keV. In this hard X-ray regime, the wind modulation is driven solely by Thomson scattering. However, at even higher energies, Klein--Nishina corrections reduce the effective scattering cross-section. Consequently, the wind's contribution to the modulation amplitude diminishes, leaving a profile governed primarily by the energy-independent geometric occultation of the central engine by the ``Turbulent Wall''. This explains the transition to a narrower minimum light curve centered on $\phi_X \sim 0$ observed in hard X-rays by \textit{INTEGRAL} and \textit{Swift}/BAT \citep{Zdziarski2012}, which traces the bulge eclipse rather than the ``Suppression zone''. In contrast, the GeV gamma-ray modulation observed by the \textit{Fermi} LAT lags significantly ($\phi_X \approx 0.3-0.4$; \citealt{Abdo2009}). As proposed by \citet{Dubus2010}, this likely arises from the eclipse of anisotropic inverse Compton scattering in a jet inclined relative to the orbital plane normal, confirming that the jet orientation is decoupled from the disk/wall geometry.

The orbital behavior of the infrared (IR) emission provides an independent test of this geometry. Historically, the IR minimum is observed to be largely synchronous with the X-ray minimum ($\phi_X \approx 0.0$) \citep{Mason1976}. In the standard wind model, this synchronization is assumed to be coincidental. If the IR modulation were driven solely by opacity variations in the global Wolf--Rayet wind, the IR minimum should coincide with the physical conjunction at $\phi_X \approx 0.11$. The fact that the IR minimum remains aligned with the ``Turbulent Wall'' ($\phi_X \approx 0.0$) strongly suggests that the Wall acts as a ``common shutter'' for both bands. This implies that the variable IR component is not diffuse, but is co-spatial with the central accretion structure---consistent with emission from the shock-heated ``turbulent'' region.

\section{Summary and Conclusions}
\label{sec:conclusions}

We have presented a quantitative geometric and kinematic resolution for the Cygnus X-3 system, reconciling the historical tension between its low inclination ($i \approx 28^\circ$) derived from \textit{IXPE} and the high-amplitude modulation of its lightcurve and emission lines. By synthesizing the \textit{RXTE}/ASM data with a ``Hybrid'' RLOF model, we arrive at the following conclusions:

\begin{enumerate}
   \item Kinetic Origin of the X-ray Minimum: The primary X-ray minimum at $\phi_X = 0.0$ is not produced by the Wolf--Rayet star itself, but by a ``Turbulent Wall''---a vertically extended, shock-heated bulge formed by the direct kinetic impact of the super-critical RLOF stream on the accretion disk rim. This kinetic structure occults the central engine approximately $\Delta\phi \sim 0.11$ prior to physical superior conjunction.
   
   \item Wind-Eclipse of the Bulge (The Iron Dip): The broad, asymmetric egress is driven by absorption in the extended Wolf--Rayet wind. This ``Suppression Zone'' begins at the true physical superior conjunction ($\phi_X \approx 0.11$) and extends to $\phi_X \approx 0.50$. During this interval, the dense stellar wind eclipses the X-ray bright ``Turbulent Wall'' located on the far side of the orbit. This geometry naturally reproduces the broad historical ``Iron Dip'' and the high-velocity absorption features detected by \textit{XRISM}.

   \item Density Stratification and the Coronagraph Effect: At the X-ray minimum the high-density Wall ($n_e \gtrsim 10^{15} \text{ cm}^{-3}$) selectively blocks the compact continuum core, leaving the lower-density ($n_e \sim 10^{12} - 10^{13} \text{ cm}^{-3}$) ambient corona/wind halo visible. This naturally explains both the observed increase in iron line equivalent width at minimum and the wind-like plasma densities derived from high-resolution X-ray spectroscopy.

   \item Origin of the Fe Orbital Modulation: The Fe\,{\sc xxvi} radial velocity zero-crossing at $\phi_X \approx 0.0$ strongly indicates the emission is associated with the turbulent stream-impact site rather than the Black Hole center of mass. The large observed velocity amplitude ($K_{obs} \approx 430$ km s$^{-1}$) arises naturally from localized stream dynamics, requiring a moderately massive black hole ($M_X \approx 15 M_{\odot}$).

   \item Non-Conservative Orbital Expansion: The observed period increase
($\dot{P}/P \approx 10^{-6}\,{\rm yr}^{-1}$; \citealt{Zdziarski2018}) arises
naturally from highly non-conservative mass transfer ($\beta\approx0$), with inner-disk
winds carrying modest specific angular momentum ($\alpha\sim1$). The same super-Eddington
transfer rate ($\approx2\times10^{-5}\,M_\odot\,{\rm yr}^{-1}$) required for the Splash Wall
geometry reproduces the expansion.
\end{enumerate}

As a nearby Galactic system with well-resolved orbital, spectral, and timing diagnostics, Cygnus X-3 thus provides a uniquely accessible laboratory for studying the geometry, dynamics, and radiative consequences of wind-focused, RLOF–like mass transfer and super-Eddington accretion from massive stellar winds—processes that are otherwise observable only in extragalactic ULXs.

\begin{acknowledgments}
I gratefully acknowledge the support of George Washington University and Professor Chryssa Kouveliotou. The constructive comments of the referees were much appreciated. The author acknowledges the use of Gemini (Google) and ChatGPT for assistance with editing and formatting this manuscript.
\end{acknowledgments}


\appendix
\section{Reproducible Analysis Code}
\label{app:code}

The following Python script reproduces the model fit, figure generation, and bootstrap error analysis presented in Section \ref{sec:modeling}.
\begin{footnotesize}
\begin{verbatim}
import numpy as np
from scipy.optimize import minimize

# --- 1. OBSERVATIONAL DATA (RXTE/ASM Folded) ---
# Data normalized to Phase 0.0 = Minimum
DATA_PHASE = np.array([
    0.05, 0.08, 0.12, 0.15, 0.18, 0.22, 0.25, 0.30, 0.35, 0.38, 
    0.42, 0.45, 0.50, 0.55, 0.60, 0.65, 0.68, 0.72, 0.75, 0.78, 
    0.82, 0.85, 0.90, 0.92, 0.96, 0.99, 1.05, 1.08, 1.12, 1.15, 
    1.18, 1.22, 1.25, 1.30, 1.35, 1.38, 1.42, 1.45, 1.50, 1.55, 
    1.60, 1.65, 1.68, 1.72, 1.75, 1.78, 1.82, 1.85, 1.90, 1.92, 
    1.96, 1.99
])
DATA_COUNTS = np.array([
    6.8, 7.5, 8.2, 8.8, 9.5, 9.8, 9.7, 10.2, 11.0, 11.8, 
    12.5, 13.5, 12.8, 13.8, 14.5, 13.5, 12.5, 13.8, 13.2, 12.5, 
    11.2, 8.5, 6.5, 6.5, 5.8, 6.2, 6.8, 7.5, 8.2, 8.8, 
    9.5, 9.8, 9.7, 10.2, 11.0, 11.8, 12.5, 13.5, 12.8, 13.8, 
    14.5, 13.5, 12.5, 13.8, 13.2, 12.5, 11.2, 8.5, 6.5, 6.5, 
    5.8, 6.2
])

# Proportional Errors: 5% of the count rate
Y_ERR = 0.05 * DATA_COUNTS

def physical_model(x_raw, p):
    """
    Synthesizes lightcurve from geometric and wind parameters.
    p = [shift, peak_ph, w_shock, decay, t_avg, t_amp, 
         min_c, max_c, H_base, H_amp, hard]
    """
    (shift, peak_ph, w_shock, decay, t_avg, t_amp, 
     min_c, max_c, H_base, H_amp, hard) = p
    
    # Transform to Physical Frame (0 = Conjunction)
    phi_phys = (x_raw + shift) % 1.0
    
    # Fixed Constants
    incl = 28.0   # Inclination (deg)
    R_disk = 1.0
    
    y_model = []
    for val in phi_phys:
        p_curr = val % 1.0
        
        # A. Splash Wall (Asymmetric Perturbation)
        diff = p_curr - peak_ph
        if diff > 0.5: diff -= 1.0
        if diff < -0.5: diff += 1.0
        
        # Ingress (Shock) vs Egress (Decay)
        if diff < 0: shape = np.exp(-0.5 * (diff / w_shock)**2)
        else: shape = np.exp(-1.0 * (diff / decay))
            
        H_curr = H_base + H_amp * shape
        
        # Geometric Visibility (Sigmoid)
        theta = np.degrees(np.arctan2(R_disk, H_curr))
        vis_geo = 1.0 / (1.0 + np.exp(-(theta - incl) * hard))
        
        # B. Wind Absorption (Cosine modulation)
        tau = t_avg + t_amp * np.cos(2 * np.pi * p_curr)
        trans_wind = np.exp(-tau)
        
        # Total Flux
        flux = (min_c + (max_c - min_c) * vis_geo) * trans_wind
        y_model.append(flux)
        
    return np.array(y_model)

def objective(p, x, y, err):
    return np.sum(((y - physical_model(x, p)) / err)**2)

# --- BOOTSTRAP ANALYSIS ---
# Initial Guesses (Updated for 5% Error Fit)
p0 = [-0.11, 0.79, 0.11, 0.29, 0.31, 0.11, 7.5, 17.7, 1.30, 1.06, 0.54]

# Bounds
bounds = [
    (-0.25, -0.05), (0.75, 0.90), (0.05, 0.20), (0.1, 0.5), 
    (0.1, 1.0), (0.0, 0.5), (5, 12), (14, 20), 
    (0.5, 2.0), (0.5, 3.0), (0.1, 5.0)
]

# Calculate Reference Residuals from Best Fit
res_main = minimize(objective, p0, args=(DATA_PHASE, DATA_COUNTS, Y_ERR), 
                    method='L-BFGS-B', bounds=bounds)
p_best = res_main.x
y_best = physical_model(DATA_PHASE, p_best)
residuals = DATA_COUNTS - y_best

print(f"Main Fit Chi2: {res_main.fun:.2f}")

print("Running Bootstrap (N=100)...")
np.random.seed(42)
boot_results = []
for i in range(100):
    res_sample = np.random.choice(residuals, size=len(residuals), replace=True)
    y_boot = y_best + res_sample
    res = minimize(objective, p_best, args=(DATA_PHASE, y_boot, Y_ERR), 
                   method='L-BFGS-B', bounds=bounds)
    if res.success: boot_results.append(res.x)

# Output Results
means = np.mean(boot_results, axis=0)
stds = np.std(boot_results, axis=0)
names = ['Shift', 'Wall_Ph', 'Shock_W', 'Decay', 'Tau_Avg', 'Tau_Amp', 
         'Min_C', 'Max_C', 'H_Base', 'H_Amp', 'Hard']

for n, m, s in zip(names, means, stds):
    print(f"{n:10s}: {m:.3f} +/- {s:.3f}")
\end{verbatim}
\end{footnotesize}

\newpage
\section{Orbital Expansion from Mass Loss and Angular-Momentum Loss}
\label{sec:Appendix_Orbital}

\subsection{Mass-Transfer Rate in the Super-Critical Regime}

The mass-transfer rate required to sustain the observed X-ray luminosity
($L_X \approx 10^{39}\,{\rm erg\,s^{-1}}$) follows from standard accretion
energetics:
\begin{equation}
\dot{M}_{\rm transfer}
=
\frac{L_X}{\beta \eta c^2},
\label{eq:mdot_req}
\end{equation}
where $\eta \approx 0.1$ is the radiative efficiency. Because Cygnus~X-3 operates
well above the Eddington limit, the compact object can accrete only a small
fraction of the supplied mass. Adopting a representative accretion efficiency
$\beta \approx 0.01$ yields a required transfer rate
$\dot{M}_{\rm transfer} \approx 2 \times 10^{-5}\,M_{\odot}\,{\rm yr^{-1}}$, consistent
with super-critical, wind-focused Roche-lobe overflow from a Wolf--Rayet donor.

\subsection{Angular-Momentum Loss and Orbital Expansion}

\begin{figure*}[t]
\centering
\includegraphics[width=\textwidth]{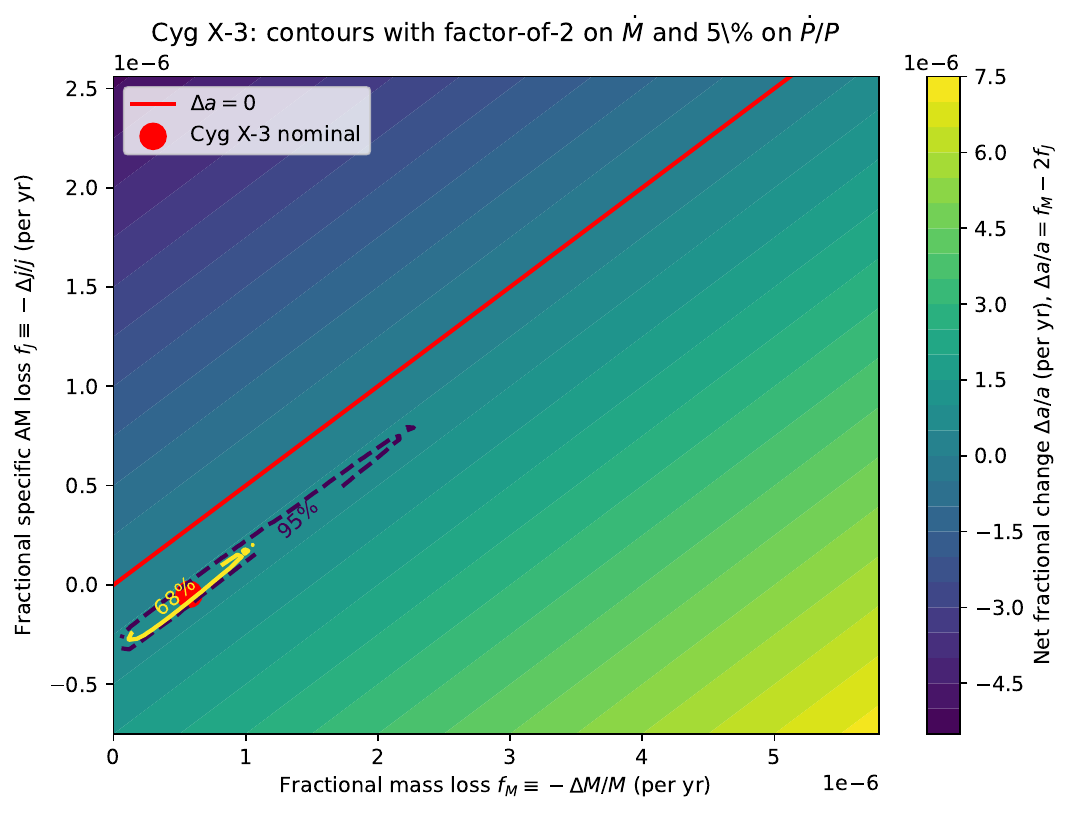}
\caption{
Orbital evolution in the plane of fractional mass loss
$f_M\equiv-\Delta M/M$ and fractional specific angular-momentum loss
$f_J\equiv-\Delta j/j$, using the scaling $\Delta a/a=f_M-2f_J$.
Colors indicate the net fractional change in orbital separation, with the red line
marking $\Delta a=0$. Solid and dashed curves show the 68\% and 95\% credible regions
for Cygnus~X-3, assuming $M_{\rm BH}=15\,M_\odot$ and $M_{\rm WR}=20\,M_\odot$, a
conservative factor-of-two uncertainty on $\dot M$, and a 5\% uncertainty on the
measured $\dot P/P$. The red point denotes the nominal values. The system lies
robustly in the $\Delta a>0$ region, demonstrating that orbital expansion is dominated
by mass loss rather than angular-momentum removal.
}
\label{fig:orbit_contours}
\end{figure*}

Following the standard non-conservative binary evolution formalism
\citep[e.g.,][]{Soberman1997,Zdziarski2018}, the orbital period evolution may be written as

\begin{equation}
\frac{\dot{P}}{P}
=
-\frac{\dot{M}_{\rm WR}}{M_{\rm WR}}
\left[
3
- 3\beta\frac{M_{\rm WR}}{M_X}
- \frac{M_{\rm WR}}{M_{\rm tot}}
- 3\alpha(1-\beta)\frac{M_X}{M_{\rm tot}}
\right],
\label{eq:pdot}
\end{equation}
where $\beta$ is the fraction of transferred mass accreted by the compact object,
$\alpha$ parameterizes the specific angular momentum carried away by the expelled
material, and $M_{\rm tot}=M_{\rm WR}+M_X$. We adopt representative component masses
$M_{\rm WR}=20\,M_\odot$ and $M_X=15\,M_\odot$, broadly consistent with current
constraints on Cygnus~X-3, and use these values in evaluating
Equations~\ref{eq:pdot}--\ref{eq:pdot_beta0} and Figure~\ref{fig:orbit_contours}.
Modest variations about these values do not affect the qualitative
result that Cygnus~X-3 lies deep in the mass-loss--dominated expansion regime.

In the extreme non-conservative limit ($\beta \rightarrow 0$),
Equation~\ref{eq:pdot} reduces to
\begin{equation}
\frac{\dot{P}}{P}
=
-\frac{\dot{M}_{\rm WR}}{M_{\rm WR}}
\left[
3
- \frac{M_{\rm WR}}{M_{\rm tot}}
- 3\alpha \frac{M_X}{M_{\rm tot}}
\right],
\label{eq:pdot_beta0}
\end{equation}

\subsection{Consistency with the Observed Period Derivative}

Adopting $\beta \approx 0$ and $\alpha \approx 1$, Equation~\ref{eq:pdot_beta0}
yields
\begin{equation}
\frac{\dot{P}}{P}
\approx
\frac{\dot M_{\rm loss}}{M_{\rm WR}}
\sim 10^{-6}\,{\rm yr^{-1}},
\end{equation}
where $\dot M_{\rm loss}\equiv-\dot M_{\rm WR}>0$ denotes the Wolf--Rayet mass-loss
rate. For $\dot M_{\rm loss}\approx2\times10^{-5}\,M_\odot\,{\rm yr^{-1}}$, this is in excellent
agreement with the observed secular expansion of the orbit
($\dot P/P \simeq 1\times10^{-6}\,{\rm yr^{-1}}$; \citealt{Singh2002,Zdziarski2018}).
This demonstrates that the measured $\dot{P}>0$ is fully compatible with super-critical
Roche-lobe overflow, provided that the mass transfer is highly non-conservative
and that the dominant mass-loss channel carries only modest specific angular
momentum. We therefore conclude that the observed orbital evolution of Cygnus~X-3 does not
argue against Roche-lobe overflow, but instead provides independent support for a
super-critical, non-conservative mass-transfer regime consistent with the
wind-focused RLOF geometry developed in this work.

\subsection{Illustration of the $(f_M,f_J)$ Plane}

To illustrate why mass loss dominates the secular orbital evolution in Cygnus~X-3, we
consider the back-of-the-envelope scaling for Keplerian orbits,
\begin{equation}
a \sim \frac{j^2}{GM},
\end{equation}
where $a$ is the orbital separation, $j$ the specific orbital angular momentum, and $M$
the total gravitating mass. Taking logarithmic derivatives gives
\begin{equation}
\frac{\Delta a}{a} = 2\,\frac{\Delta j}{j} - \frac{\Delta M}{M}.
\end{equation}
Although $a\propto j^2$, reducing $j$ requires efficient outward transport of angular
momentum prior to its removal, whereas reducing $M$ simply ejects binding mass from the
system. Consequently, even modest mass loss directly weakens the gravitational potential
and widens the orbit, while comparable changes driven by angular-momentum loss demand
large-scale angular-momentum redistribution. In essence, mass loss is a local process,
whereas angular-momentum loss is intrinsically non-local, requiring global redistribution
before escape.

Figure~\ref{fig:orbit_contours} visualizes this relation in the $(f_M,f_J)$ plane, where
$f_M\equiv-\Delta M/M$ and $f_J\equiv-\Delta j/j$ denote the fractional mass and specific
angular-momentum loss rates. We fix $M_{\rm BH}=15\,M_\odot$ and $M_{\rm WR}=20\,M_\odot$,
adopted a conservative factor-of-two uncertainty on the mass-transfer rate and a
5\% uncertainty on the measured orbital period derivative. The resulting 68\% and 95\% contours lie robustly below the
$\Delta a=0$ boundary, demonstrating that Cygnus~X-3 occupies the mass-loss--dominated
expansion regime. This provides a simple, model-independent illustration that the
observed $\dot P>0$ is naturally explained by extreme non-conservative mass transfer,
consistent with super-critical accretion and inner disk winds.

In this representation, classical BHL wind-fed systems would occupy
substantially higher $f_J/f_M$ than Cygnus~X-3. While donor wind mass loss alone can drive
orbital expansion, any wind material that is captured and subsequently re-ejected from
near the compact object carries the accretor’s orbital specific angular momentum, which
is large for mass ratios $q\gtrsim1$. This shifts wind-fed systems upward in the
$(f_M,f_J)$ plane, placing them near the $\Delta a=0$ boundary or into the contraction
regime. By contrast, Cygnus~X-3 lies deep in the low-$f_J$, mass-loss--dominated expansion
region, indicating that most escaping material carries relatively little orbital angular
momentum.

This behavior arises naturally in Roche-lobe overflow feeding a supercritical inner disk,
where radiation-driven winds are launched deep in the gravitational potential well and
therefore remove modest orbital angular momentum while efficiently reducing the system’s
gravitating mass. As a result, the same inner-disk outflows required to explain the
super-Eddington luminosity and phase-resolved iron-line diagnostics also dominate the
secular orbital evolution. The location of Cygnus~X-3 in Figure~\ref{fig:orbit_contours}
thus provides an independent, geometric argument in favor of highly non-conservative
Roche-lobe overflow as the primary mass-transfer mode in the system.

Finally, the same mass-loss--dominated expansion that explains the present $\dot P>0$
also governs the long-term evolution of the system. Although orbital widening increases
the Roche-lobe radius of the Wolf--Rayet donor, WR stars undergo rapid nuclear evolution
and sustained mass loss on $\sim10^{5}$--$10^{6}$\,yr timescales \citep{Crowther2007},
naturally maintaining near-contact. Because the supercritical accretion flow expels most
of the transferred mass from the inner disk, the system self-regulates around marginal
Roche-lobe overflow. Cygnus~X-3 can therefore persist in a wind-focused RLOF state for a
substantial fraction of the Wolf--Rayet lifetime, sustaining super-Eddington accretion
while secularly expanding its orbit.

\clearpage
\section*{Erratum: The Geometric Crisis in Cygnus X-3: Limitations of Wind-Fed Accretion and the Case for Roche-Lobe Overflow}

\section*{}
In Section~3 and the caption to Figure~1, references to the superior conjunction of the Wolf--Rayet donor should read inferior conjunction of the Wolf--Rayet donor, equivalently superior conjunction of the compact object. The binary phase $\phi_{\rm bin}=0$ places the WR star on the observer-facing side of the binary, consistent with the ``WR closest approach'' description in Figure~1B. With the published rounded phase shift, $\phi_{\rm bin}=(\phi_X-0.11)\bmod 1$, X-ray minimum precedes this conjunction by approximately $0.11$ orbital cycles; the statement in Section~3.1 that it lags conjunction is reversed. These convention corrections do not alter the implemented phase transformation or light-curve function.

In Section~3 of the published article, the rim-angle definition following Equation~(5) should be $\theta_{\rm rim}(\phi)=\arctan[R_{\rm disk}/H(\phi)]$, rather than $\arctan[H(\phi)/R_{\rm disk}]$. Both $\theta_{\rm rim}$ and the inclination $i$ are measured from the disk polar axis. Appendix~A already uses the correct expression; this correction does not change the implemented light-curve function or its numerical fit results.

Appendix~A inadvertently treats the same 26 phase bins, repeated over two cycles, as independent measurements. We have repeated the fit using each bin once and replaced the exactly degenerate parameters $F_{\min}$, $F_{\max}$, and $\tau_{\rm avg}$ by the two effective normalizations $A=F_{\min}e^{-\tau_{\rm avg}}$ and $B=F_{\max}e^{-\tau_{\rm avg}}$. The physical light-curve function and original feasible parameter bounds are retained. The corrected fit gives $\chi^2=10.075$ for 16 nominal degrees of freedom ($\chi^2_\nu=0.630$). The fitted conjunction phase, $\phi_X=0.1074$, and wall phase, $\phi_{\rm wall}=0.8400$, remain consistent with the published values. The fitted curve differs from the recovered original fit by less than $0.006$ counts~s$^{-1}$.

Table~\ref{tab:corrected_fit} replaces the original Table~1. Uncertainty estimates were recalculated using 500 parametric simulations, adding independent Gaussian errors of standard deviation $0.05$ times each original bin's count rate and refitting with multiple starting points and convergence checks. The reported percentile ranges are conditional on this assumed error model and the retained parameter bounds; they are not calibrated observational confidence intervals. Several wall-structure parameters are weakly constrained and reach bounds, so symmetric errors are inappropriate. The nominal degrees of freedom and reduced $\chi^2$ are descriptive in the presence of these constraints. The fitted morphology is preserved, but the original parameter precision is not. The accompanying corrected code supersedes the fitting and uncertainty calculation in Appendix~A. The corrected code, input data, diagnostic plots, and simulation results are publicly available\footnote{\url{https://github.com/nicholasewhite/cygx3-lightcurve}; the results reported here correspond to commit \texttt{0cf1e2d}.}.

In Section~5.3, for $P=4.8$ hr, $M_{\rm WR}=15$--$20\,M_\odot$, and $M_X=5\,M_\odot$, the binary separation should be $a=3.91$--$4.21\,R_\odot$, rather than $3$--$3.5\,R_\odot$, and the donor Roche-lobe radius should be $R_L=1.86$--$2.11\,R_\odot$, rather than $1.4$--$1.7\,R_\odot$. For $M_X=15\,M_\odot$, $R_L=1.69$--$1.90\,R_\odot$; its weak dependence on accretor mass remains valid. The adopted wind-photospheric range of $1.5$--$3\,R_\odot$ spans both underfilling and overfilling of these lobes. This comparison alone does not establish Roche-lobe overflow, since an optical-depth surface in an outflow need not bound gravitationally confined material.

In Section~4, for $M_{\rm WR}=20\,M_\odot$, $P=4.8$ hr, and $i=28^\circ$, the compact object's projected orbital semi-amplitude about the binary center of mass is approximately $400$ km~s$^{-1}$ for $M_X=5\,M_\odot$ and $319$ km~s$^{-1}$ for $M_X=15\,M_\odot$, replacing the quoted values exceeding $500$ and approximately $240$ km~s$^{-1}$, respectively.

Equation~(8) additionally approximated the observed iron-line semi-amplitude as the scalar sum of the orbital and local gas contributions. For emission associated with an azimuthally localized stream-impact bulge, these contributions generally differ in phase, and their semi-amplitudes cannot simply be added or subtracted. The light-curve model constrains the bulge's location, but translating this geometry into a velocity decomposition requires specifying the local flow direction. Consequently, the inferred wall velocity of approximately $400$ km~s$^{-1}$, and its quoted value of approximately $28\%$ of the local Keplerian speed, are withdrawn.

The argument excluding a $5\,M_\odot$ black hole and requiring approximately $15\,M_\odot$ is therefore also withdrawn. The measured semi-amplitude, $430^{+150}_{-140}$ km~s$^{-1}$, does not by itself establish a preferred compact-object mass. These corrections leave unchanged the proposed association of the iron-line emission with the stream-impact bulge and its extended atmosphere, motivated by the phase-dependent line behavior and the analogy with bright-spot S-waves in cataclysmic variables. A quantitative mass constraint from this interpretation requires modelling the amplitude and phase of the combined orbital and local flow velocities.

The corrected binary dimensions and orbital velocities are not inputs to Appendix~A's normalized light-curve function and therefore do not change it.

\begin{table*}[t]
\centering
\caption{Optimized Physical Model Parameters (Corrected)}
\label{tab:corrected_fit}
\small
\setlength{\tabcolsep}{8pt}
\begin{tabular}{lccc}
\hline\hline
\textbf{Parameter} & \textbf{Symbol} & \textbf{Best fit} & \textbf{16th--84th percentiles} \\
\hline
\multicolumn{4}{c}{\textit{System Geometry}} \\
\hline
Orbital Inclination & $i$ & $28.0^\circ$ & Fixed \\
Phase offset ($\phi_X\rightarrow\phi_{\rm bin}$) & $\Delta\phi$ & $-0.1074$ & $[-0.1416,\,-0.0921]$ \\
\hline
\multicolumn{4}{c}{\textit{Turbulent Wall (Stream Impact Bulge)}} \\
\hline
Wall Peak Phase & $\phi_{\rm wall}$ & $0.8400$ & $[0.7966,\,0.8611]$ \\
Base Rim Height & $H_{\rm base}/R_{\rm disk}$ & $0.5000$ & $[0.5000,\,1.7182]$ \\
Bulge Amplitude & $H_{\rm amp}/R_{\rm disk}$ & $2.2393$ & $[0.5000,\,3.0000]$ \\
Shock Width (Ingress) & $\sigma_{\rm shock}$ & $0.1056$ & $[0.0527,\,0.1342]$ \\
Decay Scale (Egress) & $\lambda_{\rm decay}$ & $0.2628$ & $[0.1000,\,0.5000]$ \\
Source Hardness & $k$ & $0.3935$ & $[0.2973,\,1.7644]$ \\
\hline
\multicolumn{4}{c}{\textit{Wolf--Rayet Wind}} \\
\hline
Average Optical Depth & $\tau_{\rm wind,avg}$ & \multicolumn{2}{c}{Degenerate\tablenotemark{a}} \\
Wind Modulation Amplitude & $\tau_{\rm wind,amp}$ & $0.2037$ & $[0.1472,\,0.2242]$ \\
\hline
\multicolumn{4}{c}{\textit{Flux Levels (ASM Counts s$^{-1}$)}} \\
\hline
Effective Minimum (DC Halo) & $A$ & $6.5534$ & $[5.6722,\,6.9510]$ \\
Effective Maximum (Core) & $B$ & $11.3353$ & $[11.1622,\,12.2679]$ \\
\hline
\multicolumn{4}{c}{\textit{Fit Statistics}} \\
\hline
Total Chi-Squared & $\chi^2$ & $10.075$ & --- \\
Degrees of Freedom & $\nu$ & $16$ & --- \\
Reduced Chi-Squared & $\chi^2_\nu$ & $0.630$ & --- \\
\hline\hline
\end{tabular}
\tablenotetext{a}{The mean optical depth and intrinsic flux levels are exactly degenerate. Only $A=F_{\min}e^{-\tau_{\rm wind,avg}}$ and $B=F_{\max}e^{-\tau_{\rm wind,avg}}$ are fitted; both remain subject to phase-dependent wind modulation.}
\tablecomments{For fitted parameters, the final column gives the bracketed 16th--84th percentiles from 500 parametric simulations. These ranges assume independent Gaussian errors equal to 5\% of each original bin's count rate and the original parameter bounds; they are not symmetric errors, posterior intervals, or calibrated observational confidence intervals. Boundary accumulation affects several parameters. Phase quantities are in orbital cycles and $k$ is in deg$^{-1}$. The phase offset obeys $\phi_{\rm bin}=(\phi_X+\Delta\phi)\bmod 1$. Fit statistics use 26 unique bins and 10 free parameters; $\nu$ and $\chi^2_\nu$ are nominal and descriptive.}
\end{table*}


\begin{thebibliography}{}

\bibitem[Abdo et al.(2009)]{Abdo2009} Abdo, A.~A., Ackermann, M., Ajello, M., et al.\ 2009, Science, 326, 1512

\bibitem[Abramowicz et al.(1988)]{Abramowicz1988} Abramowicz, M.~A., Czerny, B., Lasota, J.~P., \& Szuszkiewicz, E.\ 1988, \apj, 332, 646

\bibitem[Aftab et al.(2019)]{Aftab2019} Aftab, N., Paul, B., \& Kretschmar, P. 2019, \apjs, 243, 29

\bibitem[Antokhin et al.(2022)]{Antokhin2022} Antokhin, I.~I., Cherepashchuk, A.~M., Antokhina, E.~A., \& Tatarnikov, A.~M.\ 2022, \mnras, 512, 1924

\bibitem[Armitage \& Livio(1998)]{Armitage1998} Armitage, P.~J., \& Livio, M.\ 1998, \apj, 493, 898

\bibitem[Bondi \& Hoyle(1944)]{Bondi1944} Bondi, H. \& Hoyle, F.\ 1944, \mnras, 104, 273

\bibitem[Choudhury et al.(2004)]{Choudhury2004} Choudhury, M., Rao, A.~R., Vadawale, S.~V., Jain, A.~K., \& Singh, N.~S.\ 2004, \aap, 420, 665

\bibitem[Crowther(2007)]{Crowther2007} Crowther, P.~A.\ 2007, \araa, 45, 177

\bibitem[D\'iaz Trigo et al.(2006)]{DiazTrigo2006} D\'iaz Trigo, M., Parmar, A. N., Boirin, L., M\'endez, M., \& Kaastra, J. S. 2006, A\&A, 445, 179

\bibitem[Dubus et al.(2010)]{Dubus2010} Dubus, G., Cerutti, B., \& Henri, G.\ 2010, \mnras, 404, L55

\bibitem[El Mellah et al.(2019)]{ElMellah2019} El Mellah, I., Sundqvist, J.~O., \& Keppens, R. 2019, \aap, 622, L3

\bibitem[Friend \& Castor(1982)]{Friend1982} Friend, D.~B., \& Castor, J.~I. 1982, \apj, 261, 293

\bibitem[Giacconi et al.(1967)]{Giacconi1967} Giacconi, R., Gorenstein, P., Gursky, H., \& Waters, J.~R.\ 1967, \apjl, 148, L119

\bibitem[Gr{\"a}fener et al.(2012)]{Grafener2012} Gr{\"a}fener, G., Owocki, S.~P., \& Vink, J.~S.\ 2012, \aap, 538, A40

\bibitem[Hadrava \& \v{C}echura(2012)]{Hadrava2012} Hadrava, P., \& \v{C}echura, J. 2012, \aap, 542, A42

\bibitem[Hamann et al.(2006)]{Hamann2006} Hamann, W.-R., Gr{\"a}fener, G., \& Liermann, A.\ 2006, \aap, 457, 1015

\bibitem[Hanson et al.(2000)]{Hanson2000} Hanson, M.~M., Still, M.~D., \& Fender, R.~P.\ 2000, \apj, 541, 308

\bibitem[Hoyle \& Lyttleton(1939)]{Hoyle1939} Hoyle, F. \& Lyttleton, R.~A.\ 1939, Proc. Camb. Philos. Soc., 35, 405

\bibitem[Jiang et al.(2014)]{Jiang2014} Jiang, Y.-F., Stone, J.~M., \& Davis, S.~W.\ 2014, \apj, 796, 106

\bibitem[Kallman et al.(2019)]{Kallman2019} Kallman, T., et al.\ 2019, \apj, 874, 51

\bibitem[Kitamoto et al.(1994)]{Kitamoto1994} Kitamoto, S., Kawashima, K., Negoro, H., et al.\ 1994, \pasj, 46, L105

\bibitem[Koljonen \& Maccarone(2017)]{Koljonen2017} Koljonen, K.~I.~I., \& Maccarone, T.~J.\ 2017, \mnras, 472, 2181

\bibitem[Langer(1989)]{Langer1989} Langer, N.\ 1989, \aap, 220, 135

\bibitem[Liu et al.(2025)]{Liu2025} Liu, M., et al.\ 2025, \mnras, 539, 69

\bibitem[Lubow \& Shu(1975)]{Lubow1975} Lubow, S.~H., \& Shu, F.~H.\ 1975, \apj, 198, 383

\bibitem[Lubow \& Shu(1976)]{Lubow1976} Lubow, S.~H., \& Shu, F.~H.\ 1976, \apjl, 207, L53

\bibitem[Marsh(1988)]{Marsh1988} Marsh, T.~R. 1988, \mnras, 231, 1117

\bibitem[Mason et al.(1976)]{Mason1976} Mason, K.~O., Becklin, E.~E., Blankenship, L., et al.\ 1976, \apj, 207, 78

\bibitem[Mikušincová et al.(2025)]{Mikusincova2025} Mikušincová, R., et al.\ 2025, arXiv:2512.12879

\bibitem[Miura et al.(2025)]{Miura2025} Miura, D., et al.\ (\textit{XRISM} Collaboration) 2025, \pasj, 77, S86

\bibitem[Ohsuga \& Mineshige(2011)]{Ohsuga2011} Ohsuga, K., \& Mineshige, S.\ 2011, \apj, 736, 2

\bibitem[Okazaki \& Russell(2014)]{Okazaki2014} Okazaki, A.~T., \& Russell, C.~M.~P. 2014, in \textit{Suzaku--MAXI 2014: Expanding the Frontiers of the X-ray Universe}, 343, arXiv:1405.4808

\bibitem[Papaloizou \& Pringle(1977)]{Papaloizou1977} Papaloizou, J.~C.~B., \& Pringle, J.~E. 1977, \mnras, 181, 441

\bibitem[Parmar et al.(1986)]{Parmar1986} Parmar A.~N., White N.~E., Giommi P., Gottwald M., 1986, ApJ, 308, 199

\bibitem[Parsignault et al.(1972)]{Parsignault1972} Parsignault, D.~R., Gursky, H., Kellogg, E.~M., et al.\ 1972, Nature Physical Science, 239, 123

\bibitem[Poutanen et al.(2007)]{Poutanen2007} Poutanen, J., et al.\ 2007, \mnras, 377, 1187

\bibitem[Pringle(1974)]{Pringle1974} Pringle, J.~E.\ 1974, \nat, 247, 21

\bibitem[Savonije(1978)]{Savonije1978} Savonije, G.~J.\ 1978, \aap, 62, 317

\bibitem[S{\c{a}}dowski et al.(2014)]{Sadowski2014} S{\c{a}}dowski, A., Narayan, R., McKinney, J.~C., \& Tchekhovskoy, A.\ 2014, \mnras, 439, 503

\bibitem[Serlemitsos et al.(1975)]{Serlemitsos1975} Serlemitsos, P.~J., Boldt, E.~A., Holt, S.~S., Rothschild, R.~E., \& Saba, J.~L.~R.\ 1975, \apjl, 201, L9

\bibitem[Singh et al.(2002)]{Singh2002} Singh, N.~S., Rao, A.~R., \& Agrawal, P.~C.\ 2002, \aap, 392, 161

\bibitem[Soberman et al.(1997)]{Soberman1997} Soberman, G.~E., Phinney, E.~S., \& van den Heuvel, E.~P.~J.\ 1997, \aap, 327, 620

\bibitem[Szostek \& Zdziarski(2008)]{Szostek2008} Szostek, A., \& Zdziarski, A.~A.\ 2008, \mnras, 386, 593

\bibitem[Tauris \& van den Heuvel(2006)]{Tauris2006} Tauris, T.~M. \& van den Heuvel, E.~P.~J.\ 2006, in Compact Stellar X-ray Sources, ed. W. Lewin \& M. van der Klis (Cambridge: Cambridge Univ. Press), 623

\bibitem[van den Heuvel et al.(2017)]{VanDenHeuvel2017} van den Heuvel, E.~P.~J., Portegies Zwart, S.~F., \& de Mink, S.~E.\ 2017, \mnras, 471, 4256

\bibitem[van der Klis \& Bonnet-Bidaud(1989)]{vanderKlis1989} van der Klis, M., \& Bonnet-Bidaud, J.~M.\ 1989, \aap, 214, 203

\bibitem[van Kerkwijk et al.(1996)]{vanKerkwijk1996} van Kerkwijk, M.~H., Geballe, T.~R., King, D.~L., van der Klis, M., \& van Paradijs, J.\ 1996, \aap, 314, 521

\bibitem[Veledina et al.(2024a)]{Veledina2024a} Veledina, A., Muleri, F., Poutanen, J., et al.\ 2024a, Nature Astronomy, 8, 1031

\bibitem[Veledina et al.(2024b)]{Veledina2024b} Veledina, A., Poutanen, J., Bocharova, A., et al.\ 2024b, \aap, 688, L11

\bibitem[Vilhu et al.(2021)]{Vilhu2021} Vilhu, O., Kallman, T.~R., Koljonen, K.~I.~I., \& Hannikainen, D.~C.\ 2021, \aap, 649, A176

\bibitem[Vilhu \& Koljonen(2025)]{Vilhu2025} Vilhu, O., \& Koljonen, K.~I.~I.\ 2025, \aap, 699, A270

\bibitem[Waltman et al.(1994)]{Waltman1994} Waltman, E.~B., et al.\ 1994, \aj, 108, 179

\bibitem[Waltman et al.(1996)]{Waltman1996} Waltman, E.~B., et al.\ 1996, \aj, 112, 2690

\bibitem[Warner(1995)]{Warner1995} Warner, B. 1995, \textit{Cataclysmic Variable Stars} (Cambridge: Cambridge Univ. Press)

\bibitem[Weisskopf et al.(2022)]{Weisskopf2022} Weisskopf, M.~C., Soffitta, P., Baldini, L., et al.\ 2022, Journal of Astronomical Telescopes, Instruments, and Systems, 8, 026002

\bibitem[White \& Holt(1982)]{White1982} White, N.~E., \& Holt, S.~S.\ 1982, \apj, 257, 318

\bibitem[White et al.(1995)]{White1995} White, N.~E., Nagase, F., \& Parmar, A.~N.\ 1995, in X-ray Binaries, ed. W.~H.~G. Lewin, J. van Paradijs, \& E.~P.~J. van den Heuvel (Cambridge: Cambridge Univ. Press), 1

\bibitem[\textit{XRISM} Collaboration(2024)]{XRISM2024} \textit{XRISM} Collaboration 2024, \apjl, 977, L34

\bibitem[Zdziarski et al.(2012)]{Zdziarski2012} Zdziarski, A.~A., Maitra, C., Frankowski, A., et al.\ 2012, \mnras, 426, 1031

\bibitem[Zdziarski et al.(2013)]{Zdziarski2013} Zdziarski, A.~A., Miko{\l}ajewska, J., \& Belczynski, K.\ 2013, \mnras, 429, L104

\bibitem[Zdziarski et al.(2018)]{Zdziarski2018} Zdziarski, A.~A., et al.\ 2018, \mnras, 479, 4399

\end{thebibliography}
\end{document}